\documentclass[11pt,a4paper]{article}

\usepackage{amsmath}
\usepackage{amsfonts}
\usepackage{color}
\usepackage{cite}
\usepackage{graphicx}
\usepackage{float}
\usepackage{subfigure}

\title{\textbf{Defects composed of kinks and Q-balls: analytical solutions and stability}}

\author{A. Alonso-Izquierdo$^{(a,b)}$ and C. Garz\'on S\'anchez$^{(a)}$
\\ [1ex]
$^{(a)}$ Departamento de Matematica Aplicada, University of Salamanca, \\ Casas del Parque 2, 37008 - Salamanca, Spain \\ [1ex]
$^{(b)}$  IUFFyM, University of Salamanca, \\ Plaza de la Merced 1, 37008 - Salamanca, Spain \\ [1ex]}

\date{\today}

\begin{document}

\maketitle

\begin{abstract}

In this paper all the defect-type solutions in a family of scalar field theories with a real and a complex field in (1+1) dimensional Minkowski spacetime have been analytically identified. Three types of solutions have been found: (a) topological kinks without the presence of $Q$-balls, (b) defects which consist of a topological kink coupled with a $Q$-ball and (c) a one-parameter family of solutions where a $Q$-ball is combined with a non-topological soliton. The properties of these solutions and its linear stability are also discussed. 
\end{abstract}

\section{Introduction}

$Q$-balls are time-dependent non-topological solitons arising in nonlinear field theories which, in addition, conserve a Noether charge associated with a global $U(1)$ symmetry \cite{Shnir2018, Lee1992}. The physical interest of these solutions comes from the fact that $Q$-balls can be produced in the early universe in supersymmetric extensions of the standard model in such a way that they can contribute to dark matter by means of the Affleck-Dine mechanism \cite{Kusenko1998, Dine2003}. In the pioneering paper \cite{Friedberg1976}, Friedberg, Lee and Sirlin investigated the presence of this class of solutions in a theoretical model involving a complex scalar field coupled to a real scalar field in three space dimensions. The nonlinear couplings between the fields arising in this model are characterized by a quartic polynomial which means that the theory is renormalizable. The authors describe the $Q$-balls present in the model and provide a thorough scheme to analyze the linear stability of these solutions when small fluctuations that maintain the conserved Noether charge constant are applied. Theorem 3 stated in this paper establishes that the necessary and sufficient conditions to guarantee the classical stability of $Q$-balls are that the small fluctuation operator evaluated on these solutions has only one negative eigenvalue and that the derivative of the Noether charge $Q$ with respect to the internal rotation frequency $\omega$ is negative. In this prescription the frequency $\omega$ is chosen as positive. After this seminal work, the existence of $Q$-balls and its properties have been studied in different contexts, see \cite{Tsumagari2008} and references therein. Some of these particular scenarios involve complex scalar field theories \cite{Coleman1985, Coleman1986,Kusenko1997, Nugaev2014}, Abelian gauge theories \cite{Lee1989, Anagnostopolos2001, Li2001,Panin2019}, Chern-Simons theories \cite{Jackiw1990,Hong1990}, non-Abelian theories \cite{Safian1988, Safian1988b, Axenides1998}, etc. 

In general, models involving $Q$-balls are so complicated that it is not possible to obtain analytical expressions for these non-topological solitons. In recent works \cite{Bazeia2016, Bazeia2016b} $Q$-balls have been exactly calculated for some theories with one complex scalar field in (1+1)-dimensions. In this paper we address the study of a one-parameter family of field theories in (1+1) dimensions which involves the coupling between a real and a complex field. The model parameter can be understood as a measure of the deformation of the model with respect to a $O(3)$ invariant linear Sigma model. Remarkably, all the defect type solutions can be analytically identified, which makes easier the study of its properties. There exist three different types of these solutions. Firstly, a standard topological kink living in the real field component emerges without the presence of $Q$-balls. The second class can be described as defects consisting of one topological kink defined in the real component and one $Q$-ball spinning along the complex field axis. From our point of view this is a new type of solutions endowed with novel properties. For example, this coupling between a topological kink and a $Q$-ball determines a new scenario which seems to elude the applicability of the previously mentioned Theorem 3. These solutions do not verify any of the hypotheses introduced in Theorem 3. Despite this fact, they are stable as it will be proved in this paper. Finally, a one-parametric family of defects involving the presence of a non-topological soliton together to a $Q$-ball is also identified. In this case Theorem 3 can be applied to demonstrate that these solutions are unstable. 

The organization of this paper is as follows: the family of deformed $O(3)$ linear sigma models addressed in this work and its properties are introduced in Section 2. The previously mentioned defects composed of kinks and $Q$-balls are analytically identified and described in Section 3. Section 4 is dedicated to investigate the linear stability of these composite solitons. Finally, the conclusions of this work are summarized in Section 5.

\section{The model} \label{Model}

\begin{figure}[h]
	\centerline{\includegraphics[height=3.cm]{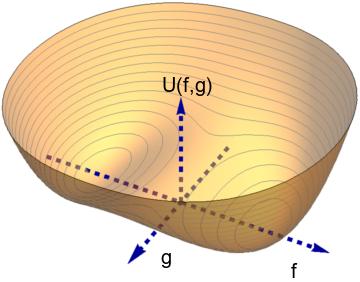} \hspace{0.5cm}\includegraphics[height=3.cm]{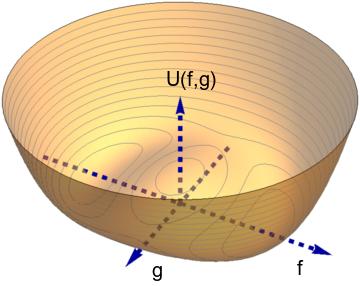} \hspace{0.5cm} \includegraphics[height=3.cm]{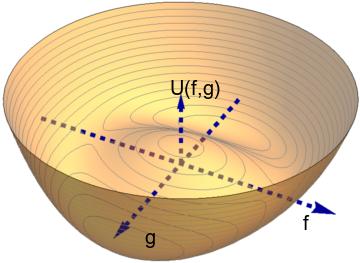}}
	\caption{\small Graphics of the effective potential $\overline{U}(f,g;\sigma)$ for the model parameter $\sigma=1.5$ and several values of the internal rotation frequency: (a) $\omega=0$, (b) $\omega=1.2$ and (c) $\omega=1.8$.} \label{fig:potential}
\end{figure}

We shall deal with a field theory immersed in a (1+1) dimensional Minkowski spacetime which involves the couplig between one real and one complex scalar field. The dynamics of this model is characterized by the action functional
\begin{equation}
S=\int d^2 x \Big[ \frac{1}{2} \partial_\mu \phi \, \partial^\mu \phi +  \frac{1}{2} \partial_\mu \overline{\psi} \, \partial^\mu \psi - U(\phi,|\psi|) \Big] \hspace{0.4cm}, \label{action}
\end{equation}
where $\phi$ and $\psi=\psi_1+i\psi_2$ are, respectively, the real and the complex scalar fields, that is, $\phi \in {\rm Maps} (\mathbb{R}^{1,1}, \mathbb{R} )$ and $\psi \in {\rm Maps} (\mathbb{R}^{1,1}, \mathbb{C} )$. In (\ref{action}) $\overline{\psi}$ stands for the complex conjugate of $\psi$. As usual in this context the Minkowski metric $g_{\mu\nu}$ is chosen as $g_{00}=-g_{11}=1$ and $g_{12}=g_{21}=0$.  The potential term $U(\phi,|\psi|)$ which will be investigated in this paper is given by the positive semi-definite expression
\begin{equation}
U(\phi,|\psi|; \sigma) = \frac{1}{2} \Big( \phi^2 + |\psi|^2 -1 \Big)^2 + \frac{1}{2} \sigma^2 |\psi|^2  \label{potential}
\end{equation}
with $\sigma\in \mathbb{R}$. The relation (\ref{potential}) is a quartic polynomial in the real field $\phi$ and the modulus of the complex field $\psi$. Note that for $\sigma=0$
\begin{equation}
U(\phi,|\psi|; 0) = \frac{1}{2} \Big( \phi^2 + \psi_1^2 + \psi_2^2 -1 \Big)^2   \hspace{0.4cm}, \label{potential0}
\end{equation}
and, therefore, this system can be understood as a deformation of a $O(3)$ linear sigma model, where the parameter $\sigma$ measures the asymmetry with respect to the rotationally invariant situation. The potential has two critical points at $(\phi,\psi)=v_\pm = (\pm 1,0)$, where the potential vanishes, $U(v_\pm)=0$. The Hessian matrix of (\ref{potential}) evaluated at these points is
\[
{\cal H}[v_\pm] = \left. \left( \begin{array}{cc} \frac{\partial^2 U}{\partial \phi^2} & \frac{\partial^2 U}{\partial \phi \, \partial |\psi|} \\[0.2cm] \frac{\partial^2 U}{\partial \phi \, \partial |\psi|} & \frac{\partial^2 U}{\partial |\psi|^2} \end{array} \right) \right|_{v_\pm} =  \left( \begin{array}{cc} 4  & 0 \\ 0 & \sigma^2  \end{array} \right) \hspace{0.4cm},
\]
which means that $v_\pm$ are minima of the potential, as expected. Despite the fact that the $O(3)$-symmetry associated to (\ref{potential0}) is broken for $\sigma\neq 0$ a $U(1)$-symmetry remains. Clearly, the model is invariant with respect to the global transformation $\psi \rightarrow e^{i\beta} \psi$, which leads to the conserved Noether charge
\begin{equation}
Q = \frac{1}{2i} \int \left( \, \overline{\psi} \,\partial_t\psi - \psi \, \partial_t \overline{\psi} \, \right) dx \hspace{0.4cm}. \label{charges0}
\end{equation}
The field equations obtained from the action funcional (\ref{action}) read
\begin{equation}
\frac{\partial^2\phi}{\partial t^2} - \frac{\partial^2 \phi}{\partial x^2} + \frac{\partial U(\phi,|\psi|)}{\partial \phi} = 0 \hspace{0.4cm},\hspace{0.4cm}
\frac{\partial^2\psi}{\partial t^2} - \frac{\partial^2 \psi}{\partial x^2} + \frac{\psi}{|\psi|} \frac{\partial U(\phi,|\psi|)}{\partial |\psi|} = 0 \hspace{0.4cm} .
\label{pde01}
\end{equation}
In this paper we are interested in searching for solutions which comprise a kink (defined by the real field) and a $Q$-ball (defined by the complex field). For this reason the ansatz
\begin{equation}
\phi(x,t) = f(x) \hspace{0.5cm},\hspace{0.5cm} \psi(x,t)= g(x) \, e^{i\,\omega\, t} \label{ansatz}
\end{equation}
is substituted into the field equations (\ref{pde01}). This leads to the system of ordinary differential equations
\begin{equation}
 \frac{\partial^2 f}{\partial x^2} = \frac{\partial U(f,g)}{\partial f}  
 \hspace{0.6cm},\hspace{0.6cm}
 \frac{\partial^2 g}{\partial x^2} = \frac{\partial U(f,g)}{\partial g} - \omega^2 g  \label{edo01}
\end{equation}
for the real functions $f(x)$ and $g(x)$. The quantity $\omega$ in (\ref{ansatz}) is the internal rotation frequency of the $Q$-ball. Without loss of generality we can consider that $\omega$ is positive. The potential term $U$ in (\ref{edo01}) becomes now
\begin{equation}
U(f,g; \sigma) =  \frac{1}{2} \Big( f^2 + g^2 -1 \Big)^2 + \frac{1}{2} \sigma^2 g^2  \label{potential03}
\end{equation}
while the conserved Noether charge (\ref{charges0}) is
\begin{equation}
Q= \omega  \int_{-\infty}^\infty (g(x))^2  dx \hspace{0.5cm}.  \label{charges}
\end{equation}
The energy functional $E[f,g]$ is written in this case as the integral over the space coordinate of the energy density ${\cal E}[f,g]$, i.e.,
\begin{equation}
E[f,g] = \int_{-\infty}^\infty {\cal E}[f,g] \, dx= \int_{-\infty}^\infty  dx \Big[  \frac{1}{2} \Big(\frac{\partial f}{d x} \Big)^2 +  \frac{1}{2} \Big(\frac{\partial g}{d x} \Big)^2 + \frac{1}{2} \omega^2 g ^2 + U(f,g; \sigma) \Big]  \label{energy}
\end{equation}
which implies that the solutions of the system must satisfy the following asymptotic conditions
\begin{equation}
\lim_{x\rightarrow \pm \infty} f (x)\in {\cal M} \hspace{0.5cm},\hspace{0.5cm} \lim_{x\rightarrow \pm \infty} \frac{d f}{dx} = 0  \hspace{0.5cm} , \hspace{0.5cm} \lim_{x\rightarrow \pm \infty} g(x) = \lim_{x\rightarrow \pm \infty} \frac{d g}{dx} = 0 
\label{asymptotics}
\end{equation}
in order to keep the total energy (\ref{energy}) finite. In (\ref{asymptotics}) ${\cal M}=\{-1,1\}$, the set of possible values of the real field leading to zeroes of the potential term $U(f,g)$. It is also clear from  (\ref{edo01}) that the problem involves the effective potential
\begin{equation}
\overline{U}(f,g; \sigma) =  U(f,g; \sigma ) -\frac{1}{2} \, \omega^2 \, g^2 =  \frac{1}{2} \Big( f^2 + g^2 -1 \Big)^2 + \frac{1}{2} (\sigma^2 -  \omega^2) \, g^2 = U(f,g;\Omega) \hspace{0.5cm}, \label{potential06}
\end{equation}
which has the same functional form as (\ref{potential03}) but with a new model parameter $\Omega$ defined as
\begin{equation}
\Omega^2 = \sigma^2-\omega^2 \hspace{0.5cm}. \label{OmegaGrande}
\end{equation}
Now, the equations (\ref{edo01}) can be written in the more compact form
\begin{equation}
	\frac{\partial^2 f}{\partial x^2} =   \frac{\partial \overline{U}(f,g)}{\partial f}  \hspace{0.6cm},\hspace{0.6cm}
	\frac{\partial^2 g}{\partial x^2} = \frac{\partial \overline{U}(f,g)}{\partial g} \hspace{0.4cm} . \label{edo2}
\end{equation}
The effective potential (\ref{potential06}) depends on the internal rotation frequency. In Figure \ref{fig:potential} the potential $\overline{U}(f,g;\sigma)$ has been depicted for several values of $\omega$ with a fixed value of the model parameter $\sigma$. The Hessian matrix of this effective potential evaluated on the points $v_\pm$ reads
\[
\overline{\cal H}[v_\pm] = \left. \left( \begin{array}{cc} \frac{\partial^2 \overline{U}}{\partial \phi^2} & \frac{\partial^2 \overline{U}}{\partial \phi \, \partial |\psi|} \\[0.2cm] \frac{\partial^2 \overline{U} }{\partial \phi \, \partial |\psi|} & \frac{\partial^2 \overline{U}}{\partial |\psi|^2} \end{array} \right) \right|_{v_\pm} =  \left( \begin{array}{cc} 4  & 0 \\ 0 & \sigma^2- \omega^2  \end{array} \right) \hspace{0.4cm}.
\]
This means that $v_\pm$ are absolute minima of $\overline{U}(f,g)$ for $\omega^2<\sigma^2$ but they become saddle points in other case. For this reason, a necessary condition for the existence of the topological defects (which we are interested in) is
\[
\omega^2 < \sigma^2 \hspace{0.5cm}.
\]

\noindent Solving the system (\ref{edo2}) together with the conditions (\ref{asymptotics}) is tantamount to finding solutions asymptotically beginning and ending at the vacuum points $v_\pm$ for Newton equations in which $x$ plays the role of time, the particle position is determined by $(f,g)$ and the potential energy of the particle is $V (f,g) = - U (f,g)$. Note that the differential equations (\ref{edo2}), or equivalently (\ref{edo01}), can be derived from the effective functional
\begin{equation}
\overline{E}[f,g;\sigma] = \int dx \Big[ \frac{1}{2} \Big( \frac{df}{dx}\Big)^2 + \frac{1}{2} \Big( \frac{dg}{dx} \Big)^2 + \overline{U}(f,g;\sigma) \Big] \hspace{0.5cm}.
\label{energybarra}
\end{equation}
keeping $\omega$ fixed, i.e., $\delta \overline{E}[f,g;\sigma]|_\omega=0$. Note that the following relation between the functionals (\ref{energy}) and (\ref{energybarra})
\begin{equation}
E[f,g;\sigma] = \overline{E}[f,g;\sigma] +\omega \, Q \label{Legendre01}
\end{equation}
holds. Alternatively, the equations (\ref{edo01}) can be derived as a stationary point of the functional (\ref{energy}) keeping $Q$ fixed, i.e, $(\delta E)|_Q=0$. All of this means that $\overline{E}[f,g;\sigma]$ is a Legendre transformation derived from the functional (\ref{energy}), see \cite{Friedberg1976}, which leads to the relations
\begin{equation}
\frac{d \overline{E}}{d\omega}\Big|_{E} = - Q \hspace{0.5cm} \mbox{and}\hspace{0.5cm} \frac{dE(Q)}{dQ} \Big|_{\overline{E}} =\omega \hspace{0.5cm}. \label{Legendre02}
\end{equation}

\section{Families of defects composed by kinks and $Q$-balls}

\label{sectionSol}

\begin{figure}[h]
	\centerline{\includegraphics[height=3.cm]{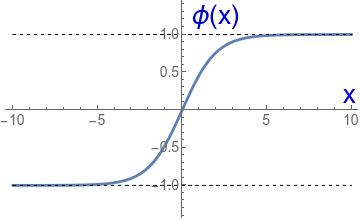}
		\hspace{1.5cm}\includegraphics[height=3.cm]{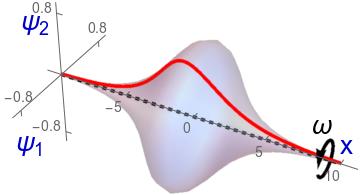} }
	\caption{\small Graphics of the solution ${\cal K}_2(x,t)$ composed by a topological kink in the real component and a $Q$-ball in the complex component of the internal space for the particular value $\Omega=0.5$ and $\alpha=0$.} \label{fig:tk2}
\end{figure}

In this Section we shall analytically identify the previously mentioned defects involving the coexistence of a kink and a $Q$-ball. The equations (\ref{edo01}) (or equivalently (\ref{edo2})) are written for our model as
\begin{equation}
\frac{\partial^2 f}{\partial x^2} =2 f (f^2+g^2-1)
\hspace{0.6cm},\hspace{0.6cm}
\frac{\partial^2 g}{\partial x^2} = 2 g (f^2+g^2-1 ) + \Omega^2 \,g \hspace{0.5cm}.  \label{edo03}
\end{equation}
These equations have been well studied in the context of multi-component kink solutions arising in the MSTB model. A thorough summary of the history and the analytical properties of this model can be found in \cite{Alonso2019} and references therein. The key point is that the equations (\ref{edo03}) can be solved by introducing elliptic variables in the internal space $(f,g)$ whose iso-coordinate curves consist of ellipses and hyperbolas with foci $F_\pm=(\pm \Omega, 0)$. In these variables the differential equations (\ref{edo03}) are separable. It can be checked that for $\Omega^2= \sigma^2 - \omega^2 \geq 1$ only a topological kink and its antikink arise and there is no room for $Q$-balls in the system. These topological defects can be expressed as
\begin{equation}
	{\cal K}_1(x)=((-1)^\alpha \, \tanh \overline{x} \, , \, 0) \hspace{0.3cm},\hspace{0.3cm} \alpha=0,1, \label{tk1}
\end{equation}
which carry a total energy $E[{\cal K}_1(x)]=4/3$. In (\ref{tk1}) $\overline{x}=x-x_0$ where $x_0$ can be interpreted as the center of the solution. Note that the multi-component notation $(f,e^{i\omega t} g)$ has been employed in (\ref{tk1}) to write the solutions. Therefore, if the second component is zero the solution does not involve $Q$-balls. Note that
\begin{equation}
Q[{\cal K}_1(x)] = 0 \hspace{0.5cm} \label{cargatk1}
\end{equation}
On the other hand, for the regime $0<\Omega^2 < 1$ the presence of $Q$ balls is possible. This implies that the necessary and sufficient condition for the existence of defects consisting of kinks and $Q$-balls in the model (\ref{potential}) is
\begin{equation}
\max\{0,\sigma^2-1\} < \omega^2 < \sigma^2 \label{condition} \hspace{0.5cm}.
\end{equation}
The previously described solutions are simply given by the expression
\begin{equation}
{\cal K}_2(x,t) = \left((-1)^\alpha \tanh (\Omega \, \overline{x}) \, , \, e^{i \omega t} \sqrt{1-\Omega^2} \, {\rm sech} \, (\Omega \, \overline{x}) \right)\hspace{0.3cm},\hspace{0.3cm} \alpha =0,1, \label{tk2}
\end{equation}
which has been illustrated in Figure \ref{fig:tk2}. It can be observed that the solution consist of a kink profile in the real field axis and a non-topological soliton ($Q$-ball) spinning in the complex component of the internal space with rotational frequency $\omega$. 

\noindent For the sake of completeness, the Noether charge (\ref{charges}) for this type of defects is
\begin{equation}
Q[ {\cal K}_2(x,t) ] =  \frac{2\omega (1-\Omega^2)}{\Omega} = \, 2 \omega \, \frac{ 1-\sigma^2 + \omega^2}{\sqrt{\sigma^2-\omega^2}} \label{cargatk2}
\end{equation}
while its total energy follows the form
\begin{equation}
E[ {\cal K}_2(x,t) ]  = \frac{2(2\omega^4-\sigma^4-\sigma^2(\omega^2-3))}{3\sqrt{\sigma^2 -\omega^2}} \hspace{0.5cm}. \label{energiatk2}
\end{equation}
The previous expressions are restricted to the range $\omega^2 \in (\sigma^2-1,\sigma^2)$ where the solutions (\ref{tk2}) are well defined. Note that $\frac{dQ[{\cal K}_2(x,t)]}{d\omega} =\Omega^{-3} [ 2\sigma^2(1-\Omega^2)+4 \omega^2 \Omega^2 ] >0$. In the usual models found in the literature this condition implies that the $Q$-balls are unstable, as stated by Theorem 3 in  \cite{Friedberg1976}. However, as it will be proved in the next Section the solutions (\ref{tk2}) elude the hypotheses of this theorem and, indeed, they are stable against small fluctuations which preserve the Noether charge (\ref{cargatk2}). It seems that the topological nature of the kink living in the real component protects the $Q$-ball constituent from decaying into the vacuum configuration. From our point of view this behavior turns the composite defects (\ref{tk2}) into a new type of solution in this context.

\begin{figure}[h]
	\centerline{\includegraphics[height=3.cm]{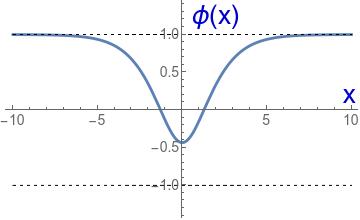}
		\hspace{0.5cm}\includegraphics[height=3.cm]{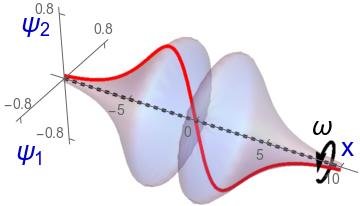} }
	\caption{\small Graphics of the solution ${\cal K}_3(x,t;\gamma)$ composed by a non-topological soliton in the real component and a $Q$-ball in the complex component of the internal space for the particular values $\sigma=0.5$, $\omega=0.25$, $\alpha=0$ and $\gamma=0$. } \label{fig:ntk}
\end{figure}

\begin{figure}[h]
	\centerline{\includegraphics[height=3.cm]{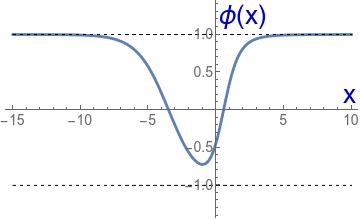}
		\hspace{0.5cm}\includegraphics[height=3.cm]{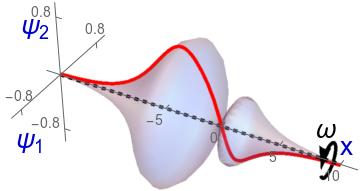} }
	\caption{\small Graphics of the solution ${\cal K}_3(x,t;\gamma)$ composed by a non-topological soliton in the real component and a $Q$-ball in the complex component of the internal space for the particular values $\sigma=0.5$, $\omega=0.25$, $\alpha=0$ and $\gamma=3$.} \label{fig:ntkb}
\end{figure}

In addition to the solutions (\ref{tk1}) and (\ref{tk2}) there exists a one-parametric family of solutions, which turn out to be a combination between a non-topological kink and a $Q$-ball. It can be checked that the expression
\begin{equation}
{\cal K}_3(x,t;\gamma) = \Big( (-1)^\alpha \frac{\Omega_- \cosh (\Omega_+ x_+) - \Omega_+ \cosh(\Omega_- x_-)}{\Omega_- \cosh (\Omega_+ x_+) + \Omega_+ \cosh(\Omega_-x_-)}, \frac{2 \Omega_+ \Omega_- e^{i\,\omega\, t}\sinh \overline{x} }{\Omega_- \cosh (\Omega_+ x_+) + \Omega_+ \cosh(\Omega_-x_-)} \Big) \label{tk3}
\end{equation}
with $\Omega_\pm = 1 \pm \Omega$, $x_\pm = \overline{x} - \gamma \,\Omega(\Omega\mp 1)$ and $\alpha=0,1$ satisfies the field equations (\ref{pde01}). Every member of the ${\cal K}_3(x,t;\gamma)$-family is determined by the value of the parameter $\gamma\in \mathbb{R}$. All of them are characterized by the presence of a non-topological kink in the real component  (asymptotically beginning and ending at the same vacuum) and the appearance of a node in the $Q$-ball profile located at $\overline{x}=0$. In Figure \ref{fig:ntk} the defect ${\cal K}_3(x,t;0)$ is depicted. For this case with $\gamma=0$ the profiles of the solution are symmetric with respect to the spatial point $\overline{x}=0$. If the two partial Noether charges 
\[
Q_1 = \omega \int_{-\infty}^0 (g(\overline{x}))^2 d\overline{x} \hspace{0.5cm},\hspace{0.5cm} Q_2 = \omega \int_0^{\infty} (g(\overline{x}))^2 d\overline{x} 
\]
are defined (such that $Q=Q_1+Q_2$), it is clear that for the ${\cal K}_3(x,t;0)$-solution the relation $Q_1=Q_2=Q/2$ holds. In Figure \ref{fig:ntkb} the member of the ${\cal K}_3(x,t;\gamma)$-family with $\gamma=3$ is plotted. Now, the solution is asymmetric and the partial charges $Q_i$ are different. In this particular case the partial charge $Q_1$ is greater than $Q_2$. This behavior continues as the value of the family parameter $\gamma$ increases. Indeed, when $\gamma$ is very large the value of $Q_2$ tends to zero and $Q_1$ tends to the Noether charge of the ${\cal K}_2(x,t)$-solution. This means that the ${\cal K}_3(x,t;\gamma)$-defects can be understood as a non-linear combination of a ${\cal K}_2(x,t)$ and a ${\cal K}_1(x,t)$ solutions.

\noindent Another remarkable properties of the previously solutions are expressed as sum rules connecting the total energies and the Noether charges of the defects. All the members of the ${\cal K}_3(x,\gamma)$-family have the same Noether charge $Q$ and the same total energy $E$. In addition to this, its conserved charge $Q$ amounts to that of the ${\cal K}_2(x)$-solution while its energy is equal to the sum of the energies of the ${\cal K}_1(x)$ and ${\cal K}_2(x)$-solutions:
\begin{eqnarray}
	&& Q[{\cal K}_3(x,t;\gamma)] = Q[{\cal K}_2(x,t)] \label{sumrule01} \\[0.2cm]	
    && E[{\cal K}_3(x,t;\gamma)] = E[{\cal K}_1(x,t)] + E[{\cal K}_2(x,t)] \label{sumrule02}
\end{eqnarray}
These results can be analytically proved from the Legendre transformations introduced in Section 2. From (\ref{Legendre02}) it is clear that
\[
Q[{\cal K}_3(x,\gamma)] = - \frac{d\overline{E}[{\cal K}_3(x,\gamma)]}{d\omega}\Big|_E = - \frac{d\overline{E}[{\cal K}_1(x)]}{d\omega}\Big|_E - \frac{d\overline{E}[{\cal K}_2(x)]}{d\omega}\Big|_E = Q[{\cal K}_1(x)]  + Q[{\cal K}_2(x)] = Q[{\cal K}_2(x)]  
\]
which justifies (\ref{sumrule01}). Here, we have used that $\overline{E}[{\cal K}_3(x,\gamma)] = \overline{E}[{\cal K}_1(x)] + \overline{E}[{\cal K}_2(x)]$, which can be manifestly demonstrated by exploiting the separability of the functional $\overline{E}(f,g)$ in elliptic coordinates, see \cite{Alonso2019}. From (\ref{Legendre01}) and (\ref{sumrule01}), the relation (\ref{sumrule02}) is directly obtained. The identities (\ref{sumrule01}) and (\ref{sumrule02}) corroborate the previously mentioned interpretation of the ${\cal K}_3(x,t;\gamma)$-solutions and leads to the results
\[
Q[{\cal K}_3(x,\gamma)] = \, 2 \omega \, \frac{ 1-\sigma^2 + \omega^2}{\sqrt{\sigma^2-\omega^2}} \hspace{0.5cm} \mbox{and} \hspace{0.5cm} E[{\cal K}_3(x,\gamma)] = \frac{4}{3} + \frac{2(2\omega^4-\sigma^4-\sigma^2(\omega^2-3))}{3\sqrt{\sigma^2 -\omega^2}} 
\] 

\subsection{Stability analysis of the $Q$-balls}

In this Section the (classical) linear stability of the solutions described in Section 2 is investigated following the prescription established in the seminal paper \cite{Friedberg1976}. In this scheme a static solution ${\cal K}(x)=(f(x),g(x))$ is perturbed by applying a small fluctuation $(\delta f,\delta g)$ which conserves the Noether charge $Q$. In order to attain this condition the internal rotation frequency of the perturbed solution must be varied by the magnitude
\[
\delta \omega = -\frac{2\omega^2}{Q} \int_{-\infty}^\infty g \, \delta g \, dx \hspace{0.5cm} .
\]
Now, the effect of these fluctuations on the energy functional (\ref{energy}) is analyzed. If the total energy $E[{\cal K}(x)+(\delta f,\delta g)]$ of the perturbed configuration is less than $E[{\cal K}(x)]$ then the solution ${\cal K}(x)$ will be unstable. It can be checked that the variation of the energy functional $E$ at second order is given by  
\begin{equation}
\delta E^{(2)}|_Q =  \int_{-\infty}^\infty dx \frac{1}{2} (\delta F)^t \, {\cal H}[{\cal K}(x)]\, \delta F + \frac{2\omega^3}{Q} \Big( \int_{-\infty}^\infty g \, \delta g \, dx \Big)^2 \label{energyvar}
\end{equation}
where the compact notation $\delta F = (\delta f,\delta g)^t$ has been used. The second order small fluctuation operator ${\cal H}[{\cal K}(x)]$ arising in the previous expression reads
\begin{equation}
{\cal H} [{\cal K}(x)]= \left( \begin{array}{cc}
-\frac{d^2}{dx^2} + \left. \frac{\partial^2 U}{\partial f^2}  \right|_{{\cal K}(x)} & \left.\frac{\partial^2 U}{\partial f \partial g} \right|_{{\cal K}(x)} \\[0.3cm]
\left.\frac{\partial^2 U}{\partial f \partial g} \right|_{{\cal K}(x)}  & -\frac{d^2}{dx^2} + \left.\frac{\partial^2 U}{\partial g^2} \right|_{{\cal K}(x)}-\omega^2
\end{array}\right)   \hspace{0.5cm} . \label{operator2}
\end{equation}
In particular for our model where the field potential term $U(f,g)$ is determined by the expression (\ref{potential03}) the operator (\ref{operator2}) can be written as
\begin{equation}
{\cal H} [{\cal K}(x)]=\left. \left( \begin{array}{cc}
-\frac{d^2}{dx^2} + 2(3 f^2 + g^2 -1)  & 4 f g \\[0.3cm]
4 f g  & -\frac{d^2}{dx^2} + 2(f^2+ 3 g^2 -1) + \Omega^2  \end{array}\right) \right|_{{\cal K}(x)} \hspace{0.5cm} . \label{operator3}
\end{equation}

\begin{figure}[h]
	\centerline{\includegraphics[height=3.cm]{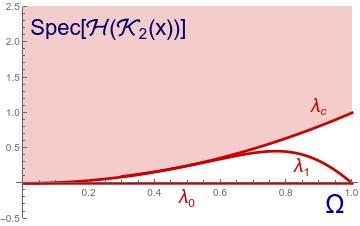}}
	\caption{\small Spectrum of the second order small fluctuation operator ${\cal H} [{\cal K}_2(x)]$ as a function of the parameter $\Omega$.} \label{fig:EspectroTK2}
\end{figure}

\noindent \textit{Linear stability analysis for the ${\cal K}_2(x)$-solutions:} To study the linear stability of the defects ${\cal K}_2(x)$, composed by a topological kink and a $Q$-ball, the expression (\ref{tk2}) is substituted into the operator (\ref{operator3}). This leads to the particular Schr\"odinger-type matrix operator
\begin{equation}
{\cal H} [{\cal K}_2(x)]= \left( \begin{array}{cc}
-\frac{d^2}{dx^2} + 4-2(2+\Omega^2)\,{\rm sech}^2  (\Omega \, x)  & 4 \sqrt{1-\Omega^2} \, {\rm sech}   (\Omega \,x )\tanh (\Omega \,x) \\[0.3cm]
4 \sqrt{1-\Omega^2} \, {\rm sech} (\Omega \, x) \tanh (\Omega\, x) & -\frac{d^2}{dx^2} + \Omega^2+2(2-3\Omega^2)\,{\rm sech}^2 (\Omega\, x) \end{array}\right) \hspace{0.5cm},   \label{operator4}
\end{equation}
which depends on the parameter $\Omega$. It can be analytically proved that $\frac{d {\cal K}_2(x)}{dx}$ is a zero mode of the operator (\ref{operator4}). However, the rest of the eigenvalues of ${\cal H} [{\cal K}_2(x)]$ must be numerically calculated. In Figure \ref{fig:EspectroTK2} the espectrum of this operator is displayed as a function of the parameter $\Omega$. It can be observed the presence of the previously mentioned zero mode with eigenvalue $\lambda_0=0$. The continuous spectra emerge on the threshold values $\Omega^2$ and $4$. Note that a discrete eigenfunction with eigenvalue $\lambda_1$ emerges for $\Omega > 0.6$ approximately. The crucial point here is that there are no negative eigenvalues. Therefore, the two contributions in (\ref{energyvar}) are positive, which means that no linear fluctuations can decrease the energy of the solution ${\cal K}_2(x)$. We have proved that this defect is stable. We recall that Theorem 3 in \cite{Friedberg1976} states that the necessary and sufficient conditions for $\delta E^{(2)}|_Q > 0$ are (i) ${\cal H}[{\cal K}(x)]$ has only one negative eigenvalue and (ii) $\frac{dQ[{\cal K}(x)]}{d\omega}<0$. However, in our case there is no negative eigenvalues of the operator ${\cal H} [{\cal K}_2(x)]$ and $\frac{dQ[{\cal K}_2(x)]}{d\omega}>0$. As a consequence, the $K_2(x)$-solutions in our model constitutes a counterexample of the universality of the previously mentioned theorem, which from our point of view always assumes the existence of negative eigenvalues.

\noindent \textit{Linear stability analysis for the ${\cal K}_3(x;\gamma)$-solutions:} The situation is more complicated for the ${\cal K}_3(x;\gamma)$-solutions. Now, the components of the fluctuation operator (\ref{operator3}) are
\begin{eqnarray*}
&& {\cal H}_{11} [{\cal K}_3(x;\gamma)] = -\frac{d^2}{dx^2} -2 + \frac{6(-\Omega_+ \cosh (\Omega_- x_-)+\Omega_- \cosh(\Omega_+ x_+))^2+ 8 \Omega_-^2 \Omega_+^2 \sinh^2 x}{(\Omega_+ \cosh (\Omega_- x_-)+\Omega_- \cosh(\Omega_+ x_+))^2}  \hspace{0.3cm} ,\\
&& {\cal H}_{12} [{\cal K}_3(x;\gamma)] = \frac{8\Omega_+ \Omega_- (\Omega_- \cosh (\Omega_+ x_+)-\Omega_+ \cosh(\Omega_- x_-)) \sinh x}{(\Omega_+ \cosh (\Omega_- x_-)+\Omega_- \cosh(\Omega_+ x_+))^2} \hspace{0.3cm} ,\\
&& {\cal H}_{22} [{\cal K}_3(x;\gamma)] = -\frac{d^2}{dx^2} -2 +\Omega^2 + \frac{2(-\Omega_+ \cosh (\Omega_- x_-)+\Omega_- \cosh(\Omega_+ x_+))^2+ 24 \Omega_-^2 \Omega_+^2 \sinh^2 x}{(\Omega_+ \cosh (\Omega_- x_-)+\Omega_- \cosh(\Omega_+ x_+))^2} \hspace{0.3cm} .
\end{eqnarray*}
Despite the intricate form of this operator some results can be formulated. For example, it can be checked that the expressions $\frac{\partial {\cal K}_3(x;\gamma)}{\partial x}$ and $\frac{\partial {\cal K}_3(x;\gamma)}{\partial \gamma}$ are zero modes of the operator ${\cal H}[{\cal K}_3(x;\gamma)]$. Another theoretical result can be derived by applying Morse theory to the space of the orbits traced by the solutions. It can be verified that all the members of the ${\cal K}_3(x;\gamma)$-family cross the point $((-1)^\alpha \Omega,0)$ (depending on the value of $\alpha$ in (\ref{tk3})). This implies the existence of a negative eigenvalue in the spectrum of the operator ${\cal H}[{\cal K}_3(x;\gamma)]$. In this scenario the hypotheses of Theorem 3 in \cite{Friedberg1976} are recovered and the claim stated there is now valid. Because these solutions verify that $\frac{dQ[{\cal K}_3(x;\gamma)]}{d\omega}>0$ this means that the solutions in the ${\cal K}_3(x;\gamma)$-family are unstable. 

We complete this stability analysis by noting that the ${\cal K}_2(x)$-solutions involve absolute stability. Again, the topological nature of the kink in the real component provides these defects with this property. A heuristic argument proving this fact is as follows. The energy of plane wave solutions around the vacua $v_\pm$ in the complex component with Noether charge $Q$ is given by $E_{\rm free} \approx \sigma Q$, see \cite{Lee1992, Tsumagari2008}. However, the topological kink in the real component cannot decay into one of the vacua. Indeed, the fact that plane waves are defined in the complex component implies that the topological kink found in this configuration must correspond to that of the ${\cal K}_1(x)$-solution (\ref{tk1}). Now, we have to compare the energy of the ${\cal K}_2(x)$-defect with that of this vibrating ${\cal K}_1(x)$-solution. It can be checked that
\[
E[ {\cal K}_2(x,t) ] < E[ {\cal K}_1(x,t) ] + \sigma Q[{\cal K}_2(x)] \hspace{0.5cm},
\]
which confirms that the ${\cal K}_2(x,t)$-defects are stable with respect to decay into free particles.

\section{Summary}

In this paper the existence of defects involving the coupling between kinks and $Q$-balls has been investigated in a one-parameter family of field theories in (1+1) dimensions with a real and a complex field. It has been found that there exist three types of solutions: ${\cal K}_1(x)$-solutions (formed by only one topological kink), ${\cal K}_2(x)$-solutions (which consist of a topological kink together with a $Q$-ball) and the one-parameter family of ${\cal K}_3(x;\gamma)$-solutions (where a $Q$-ball is combined with a non-topological soliton). All of these solutions have been analytically identified. In addition, the second of the previously mentioned solutions can be considered as a counterexample of the universality of the Theorem 3 introduced in the seminal paper \cite{Friedberg1976}. The small fluctuation operator evaluated on the ${\cal K}_2(x)$-solutions has no negative eigenvalues and the derivative of the Noether charge of these defects with respect to the frequency is positive. However, the ${\cal K}_2(x)$ solutions are stable. The topological charge of the kink living in the real component seems to prevent the $Q$-ball from decaying into the vacuum. From this point of view these solutions involve novel properties with respect to the usual $Q$-balls arising in the literature. Finally, the family of ${\cal K}_3(x;\gamma)$-defects are unstable. In this case, the previously mentioned theorem can be applied to prove this behavior.

\section*{Acknowledgments}

This research was funded by the Spanish Ministerio de Ciencia e Innovación (MCIN) with funding from the European Union NextGenerationEU (PRTRC17.I1) and the Consejería de Educación, Junta de Castilla y Le\'on, through QCAYLE project, as well as MCIN project PID2020-113406GB-I00 MTM.


\begin{thebibliography}{99}

\addcontentsline{toc}{section}{References}

\bibitem{Shnir2018}
Shnir, Y. M., \textit{Topological and non-topological solitons in scalar field theories}. Cambridge University Press, 2018.

\bibitem{Lee1992}
Lee, T.D. and Y. Pang, Phys. Rep. 221 (1992) 251.

\bibitem{Kusenko1998}
Kusenko, A. and Shaposhnikov, M.; Phys. Lett. B. 418 (1998) 46.

\bibitem{Dine2003}
Dine, M. and Kusenko, A.; Rev. Mod. Phys. 76 (2003) 1.

\bibitem{Friedberg1976}
Friedberg, R.; Lee, T.D. and Sirlin, A.; Phys. Rev. D 13 (1976) 2739.

\bibitem{Tsumagari2008}
Tsumagari, M.I.; Copeland, E.J.; Saffin, P.M.; Phys. Rev. D 78 (2008) 065021.


\bibitem{Coleman1985}
Coleman, S., Nucl. Phys. B, 262 (1985) 263.

\bibitem{Coleman1986}
Coleman, S., Nucl. Phys. B, 269 (1986) 744.

\bibitem{Kusenko1997}
Kusenko, A., Phys. Lett. B, 404 (1997) 285.

\bibitem{Nugaev2014}
Nugaev, E.Y. and Smolyakov, M.N.; JHEP, 07 (2014) 009.

\bibitem{Lee1989}
Lee, K.M.; Stein-Schabes, J.A.; Watkinks, R. and Widrow, L.M.; Phys. Rev. D39 (1989) 1665.

\bibitem{Anagnostopolos2001}
Anagnostopoulos, K.N., Axenides, M.; Floratos, E.G. and Tetradis, N.; Phys. Rev. D 64 (2001) 125006

\bibitem{Li2001}
Li, X.Z.; Hao, J.G.; Liu, D.J. and Chen, G.; J. Phys. A 34 (2001) 1459.

\bibitem{Panin2019}
Panin, A.G. and Smolyakov, M.N.; Eur. Phys. J. C, 79 (2019) 150.

\bibitem{Jackiw1990}
Jackiw, R. and Weinberg, E.J.; Phys. Rev. Lett., 64 (1990) 2234.

\bibitem{Hong1990}
Hong, J., Kim, Y. and Pac, P.Y.; Phys. Rev. Lett., 64 (1990) 2230.

\bibitem{Safian1988}
Safian, A.M.; Coleman, S.R. and Axenides, M.; Nucl. Phys. B297 (1988) 498.

\bibitem{Safian1988b}
Safian, A.M.; Nucl. Phys. B304 (1988) 392.

\bibitem{Axenides1998}
Axenides, M.; Floratos, E. and Kehagias; Phys. Lett. B 444 (1998) 190.

\bibitem{Deshaies2006}
Deshaies-Jacques, M. and MacKenzie, R.; Phys. Rev. D 74 (2006) 025006

\bibitem{Bazeia2016}
Bazeia, D.; Marques, M.A. and Menezes, R.; Eur. Phys. J. C, 76 (2016) 241.

\bibitem{Bazeia2016b}
Bazeia, D.; Losano, L.; Marques, M.A.; Menezes, R. and Da Roucha, R.; Phys. Lett. B, 758 (2016) 146.

\bibitem{Alonso2019}
Alonso-Izquierdo, A.; Physica Scripta 94 (2019) 085302.



\end{thebibliography}
\end{document}